\documentclass[twocolumn,pre,twoside,groupedaddress,floatfix,noshowpacs]{revtex4}

\usepackage{amsmath}
\usepackage{amssymb}
\usepackage{graphicx}
\usepackage{bm}
\usepackage{color}

\def\ud{\text{d}}

\begin{document}

%-------------------------------------------------------------------------------

\title{Electrolyte solutions at curved electrodes. I. Mesoscopic approach}

\author{Andreas Reindl}
\email{reindl@is.mpg.de}
\author{Markus Bier}
\email{bier@is.mpg.de}
\author{S. Dietrich}
\affiliation
{
   Max-Planck-Institut f\"ur Intelligente Systeme, 
   Heisenbergstr.\ 3,
   70569 Stuttgart,
   Germany
}
\affiliation
{
   IV. Institut f\"ur Theoretische Physik,
   Universit\"at Stuttgart,
   Pfaffenwaldring 57,
   70569 Stuttgart,
   Germany
}

\date{28 March, 2017}

\begin{abstract}
   Within the Poisson-Boltzmann (PB) approach electrolytes in contact with planar, spherical, and cylindrical electrodes are analyzed systematically.
   The dependences of their capacitance $C$ on the surface charge density $\sigma$ and the ionic strength $I$
   are examined as function of the wall curvature. The surface charge density has a strong effect on the capacitance
   for small curvatures whereas for large curvatures the behavior becomes independent of $\sigma$.
   An expansion for small curvatures gives rise to capacitance coefficients which depend only on a single parameter,
   allowing for a convenient analysis.
   The universal behavior at large curvatures can be captured by an analytic expression.
\end{abstract}

\maketitle

\section{\label{section:introduction}Introduction}
An electrical double layer capacitor or supercapacitor basically consists of electrodes which are insulated by a separator and which are in contact with an electrolyte.
Supercapacitors are used as alternative electrical energy
storage devices and combine the properties of conventional batteries, with high energy but low power densities, and conventional capacitors with the
opposite characteristics \cite{Kiamahalleh2012}.
They are used in electric vehicles and mobile phone equipments. Moreover, in search of sustainable energy systems there is still growing interest in double layer capacitors.
The capacitive behavior is determined by the nature of the electrode material, e.g., its porosity and accessible surface area. Often carbon is the electrode material of choice 
and especially ordered carbon allotropes have received much attention because their micro-texture influences the electronic properties.
Different kinds of carbon nanostructured materials,
including carbon nanotubes, carbon nanorods, spherical fullerenes, and carbon nano-onions, have been used as electrodes \cite{Kiamahalleh2012,Mombeshora2015}.
Fiber-shaped supercapacitors exhibit low weight and high flexibility and thus are promising candidates for power sources in wearable electronics \cite{Qu2016}.
In contrast to conventional capacitors with smooth electrode morphologies, supercapacitors exhibit highly curved surfaces in order to obtain large specific areas, i.e.,
high porosity. This poses the problem of understanding the properties of electric double layers at curved geometries.
A suitable method to model an electric double layer is given by the Poisson-Boltzmann (PB) theory. Within this mesoscopic approach the focus is on length scales larger than the ions
or solvent molecules because electrolyte solutions are taken to consist of pointlike
ions dissolved in a homogeneous solvent which is described by its electric permittivity only. The PB theory has been pioneered by Gouy \cite{Gouy1910} and Chapman
\cite{Chapman1913} in the 1910s and sometimes it is referred to as the Gouy-Chapman theory.
Although the model is simple, reliable predictions can be expected to hold for low ionic strengths
(below $0.2\,\text{M}=0.2\,\text{mol}/\ell$) and low electrode potentials (below $80\,\text{mV}$), in the case of aqueous solutions and monovalent salts \cite{Butt2003}.
For that reason and due to its simplicity the PB theory is used frequently.
Under certain circumstances it even allows for exact solutions, e.g., 
for electrolyte solutions at planar electrodes \cite{Butt2003,Schmickler2010}. 
Recently, exact results have been presented for an electrolyte bounded by parallel plates or inside a cylindrical charged wall if
only counterions are considered \cite{Samaj2016}. This setup might be used as a description for ions confined
in a charged nanotube or pore. However, for the corresponding spherical system, so far a solution in closed form has not been found. 
In Ref.~\cite{Samaj2015} the same authors presented an expansion for the solution of the PB equation
in spherical and cylindrical geometries with large radii of curvature, which
might resemble charged macromolecules surrounded by an electrolyte solution and which comes closest to an analytic solution of the full PB equation
for these geometries.
Within the framework of the linearized Gouy-Chapman-Stern theory in Ref.~\cite{Kant2013} a model for an arbitrary surface morphology was developed.
This facilitates, for example, the calculation of capacitances of nanostructured electrodes,
the study of which might contribute to the development of efficient energy generating and storage devices.
But also for numerical studies the PB equation often is the model of choice because its simplicity allows for fast calculations:
In order to understand the properties of the diffuse double layer at charged nanoelectrodes or carbon nanotubes
the PB equation was solved in Refs.~\cite{Dickinson2009,Henstridge2010} for spherical and cylindrical electrodes. The potential and capacitance were analyzed
for various values of the electrode radius. The evolution of capacitance models for supercapacitors gave rise to the study
in Ref.~\cite{Huang2010} in which cylindrical and slit pores were
considered within the Gouy-Chapman-Stern model to address, inter alia, the issue of how the pore shape affects the capacitance.

However, to our knowledge, so far the dependence of the capacitance on the geometry has only been addressed on a sample basis, i.e., for particular choices of system parameters.
The intention of the present work is to study the curvature dependence of the capacitance systematically within the entire, relevant parameter space.

In the present study the PB equation is solved for electrolytes surrounding spherical and cylindrical electrodes (see Sec.~\ref{section:model}).
In addition to presenting results for a variety of parameter choices, a thorough overview of the spectrum of solutions is given.
To that end the dependence of the differential capacitance on the various parameters is analyzed in detail
within the PB theory for these geometries.
We are able to discuss the limiting behavior systematically, i.e., the dependence on only one parameter or analytically.
This facilitates the understanding of the essential behavior of the data of interest which in the present case is the differential capacitance
as a function of the wall curvature.
In Sec.~\ref{subsection:lPB} a short overview of the exact results within the linearized theory is given before in Sec.~\ref{subsec:PB} the full PB equation
for various choices of the parameters is solved. In the subsequent Secs.~\ref{subsec:large_radii} and \ref{subsec:small_radii} general trends for large and small
radii of the electrodes are worked out. Corresponding technical details are discussed in Appendices \ref{app:large_radii} and \ref{app:small_radii}.
Summary and outlook are given in Sec.~\ref{sec:summary}.

\section{\label{section:model}Model}
Consider an electrolyte composed of pointlike, monovalent ions, i.e., particles without volume carrying positive or negative elementary charge $\pm e$.
Due to local charge neutrality the number densities of both ion species in the bulk are equal to the ionic strength $I$.
The solvent is regarded as a dielectric continuum with homogeneous relative permittivity $\epsilon$. The electrostatic potential $\Phi$ in this system
obeys the Poisson-Boltzmann (PB) equation \cite{Butt2003,Schmickler2010}
\begin{align}
  \Delta\Phi(\boldsymbol{r})=\frac{2eI}{\epsilon_0\epsilon}\sinh[\beta e \Phi(\boldsymbol{r})],
  \label{eq:PB}
\end{align}
where $\Delta$ is the Laplace operator, $\boldsymbol{r}\in\mathbb{R}^3$ denotes a position in three-dimensional space,
$\epsilon_0$ is the vacuum permittivity, $\beta=(k_BT)^{-1}$ with the Boltzmann constant $k_B$ and the
absolute temperature $T$. The electrolyte is assumed to be in contact with a convex electrode of planar, spherical, or cylindrical shape. The electrode is
described as a homogeneously charged hard wall with surface charge density $\sigma$. Under these assumptions
the potential $\Phi$ in Eq.~(\ref{eq:PB}) depends on a single spatial variable $\Phi(r):=\Phi(\boldsymbol{r})=\Phi(x,y,z)$ where the meaning of $r$ depends on the
geometry:
\begin{itemize}
   \item A planar wall occupies the half space $z<0$ which leads to a dependence of $\Phi$ on $r:=z$,
   \item a spherical wall $x^2+y^2+z^2<R^2$ of radius $R$ gives rise to a dependence of the potential on $r:=\sqrt{x^2+y^2+z^2}$, and for a
   \item cylindrical wall $x^2+y^2<R^2$ of radius $R$ the potential depends on $r:=\sqrt{x^2+y^2}$.
\end{itemize}
By introducing the parameter $d$ in order to distinguish the three geometries the PB equation~(\ref{eq:PB}) may be formulated in a one-dimensional fashion:
\begin{align}
   \begin{aligned}
      &\frac{1}{r^d}\frac{\partial}{\partial r}\left[r^d\frac{\partial\Phi(r)}{\partial r}\right]
                 =\frac{2eI}{\epsilon_0\epsilon}\sinh[\beta e \Phi(r)],\\
      &d=
      \begin{cases}
         0,\quad\text{planar wall},\\
         1,\quad\text{cylindrical wall},\\
         2,\quad\text{spherical wall}.
      \end{cases}
   \end{aligned}
   \label{eq:PB_1D}
\end{align}
Solutions of Eq.~(\ref{eq:PB_1D}) are subject to boundary conditions at the wall surface $r_w$ and in the bulk $r\rightarrow\infty$:
\begin{align}
   \begin{aligned}
      &\Phi'(r)\Big|_{r=r_w}=-\frac{\sigma}{\epsilon_0\epsilon},\quad r_w=
      \begin{cases}
         0,&d=0,\\
         R,&d\in\{1,2\},
      \end{cases}\\
      &\Phi'(r)\Big|_{r=\infty}=0.
   \end{aligned}
   \label{eq:BC}
\end{align}
For the considered geometries the electric field exhibits only a component $E(r)=-\Phi'(r)$
in direction normal to the wall surface. The value of the component at the surface $r_w$
is linked to the surface charge density $\sigma$ by the first boundary condition.
The second boundary condition ensures global charge neutrality. In the following we additionally demand $\Phi(\infty)=0$
so that the lower boundary
condition is fulfilled and the arbitrary integration constant of $\Phi$ is set.

For small values of the dimensionless potential $\beta e\Phi(r)\rightarrow0$ it is sufficient to consider the expansion of the hyperbolic sine in the
PB equation~(\ref{eq:PB_1D}) only up to linear order in order to obtain the linearized PB equation
\begin{align}
   \frac{1}{r^d}\frac{\partial}{\partial r}\left[r^d\frac{\partial\Phi(r)}{\partial r}\right]
                 =\kappa^2 \Phi(r),\quad
   \kappa:=\sqrt{\frac{2e^2I\beta}{\epsilon_0\epsilon}},
   \label{eq:lPB}
\end{align}
with the inverse Debye length $\kappa$.

\section{\label{section:discussion}Discussion}
The differential capacitance is defined by \cite{Schmickler2010}
\begin{align}
   C:=\frac{\partial\sigma}{\partial\Phi(r_w)}
   \label{eq:C}
\end{align}
as the change of the surface charge density $\sigma$ upon changing the potential at the wall $\Phi(r_w)$. Here the
theoretical results are presented in terms of this measurable quantity in order to facilitate comparison with experiments.
Our examinations focus on the dependence of the 
capacitance $C$ on the curvature $1/R$ of a spherical and a cylindrical wall.

\subsection{\label{subsection:lPB}Linearized PB equation}
The linearized PB equation~(\ref{eq:lPB}) can be solved analytically for the geometries under consideration:
\begin{itemize}
   \item At a planar wall the potential at the electrode is
         \begin{align}
            \Phi(0)=\frac{\sigma}{\epsilon_0\epsilon\kappa},
         \end{align}
         and the capacitance is given by the double-layer capacitance $\epsilon_0\epsilon\kappa$ \cite{Schmickler2010}
         \begin{align}
            \frac{C}{\epsilon_0\epsilon\kappa}=1.
         \end{align}
         This quantity will be used as a reference in order to define dimensionless capacitances.
   \item For a spherical wall one has
         \begin{align}
            \Phi(R)=\frac{\sigma}{\epsilon_0\epsilon\kappa}\frac{\kappa R}{\kappa R+1},
            \label{eq:PhiR_sph_lin}
         \end{align}
         and
         \begin{align}
            \frac{C}{\epsilon_0\epsilon\kappa}=1+\frac{1}{\kappa R},
            \label{eq:C_sph_lin}
         \end{align}
         a polynomial of linear order in the dimensionless curvature $(\kappa R)^{-1}$.
   \item In case of a cylindrical wall the electrode potential
         \begin{align}
            \Phi(R)=\frac{\sigma}{\epsilon_0\epsilon\kappa}\frac{\text{K}_0(\kappa R)}{\text{K}_1(\kappa R)},
            \label{eq:PhiR_cyl_lin}
         \end{align}
         and the capacitance
         \begin{align}
            \begin{aligned}
               \frac{C}{\epsilon_0\epsilon\kappa}&=\frac{\text{K}_1(\kappa R)}{\text{K}_0(\kappa R)}\\
                                                 &=1+\frac{1}{2}\frac{1}{\kappa R}-\frac{1}{8}\frac{1}{(\kappa R)^2}+O\left[\frac{1}{(\kappa R)^3}\right]
            \end{aligned}
            \label{eq:C_cyl_lin}
         \end{align}
         are given by a ratio of modified Bessel functions \cite{Abramowitz1970}. For large radii $\kappa R\gg1$ the expression can be represented by an infinite
         power series in the curvature, the truncated version of which is shown in the second line of Eq.~(\ref{eq:C_cyl_lin}).
\end{itemize}
Already the linearized PB equation reveals interesting differences for the curvature dependencies of the capacitance. Whereas in the case of the spherical wall the
entire curvature dependence is given by a linear polynomial, in the case of the cylindrical wall a transcendental expression is found.
The leading curvature correction in the case of the spherical electrode [Eq.~(\ref{eq:C_sph_lin})] is twice of that in the cylindrical case
[Eq.~(\ref{eq:C_cyl_lin})]. In the following it turns out that this ratio also holds for systems in accordance with the nonlinear PB equation
(see Sec.~\ref{subsec:large_radii}).
The analysis based on the linearized PB equation becomes rather complex when arbitrary curved surfaces are considered, which was addressed
in Ref.~\cite{Kant2013}.
However, the linear theory is valid only in the limit $\Phi\rightarrow0$ and the 
capacitances are independent of $\Phi$ or $\sigma$. Strictly speaking the full PB equation has to be considered as soon as systems with
non-vanishing $\Phi$ or $\sigma$ are of interest. This will be the focus in the following sections.

\subsection{\label{subsec:PB}Full non-linear PB equation}
\begin{figure}[!t]
   \includegraphics[width=0.44\textwidth]{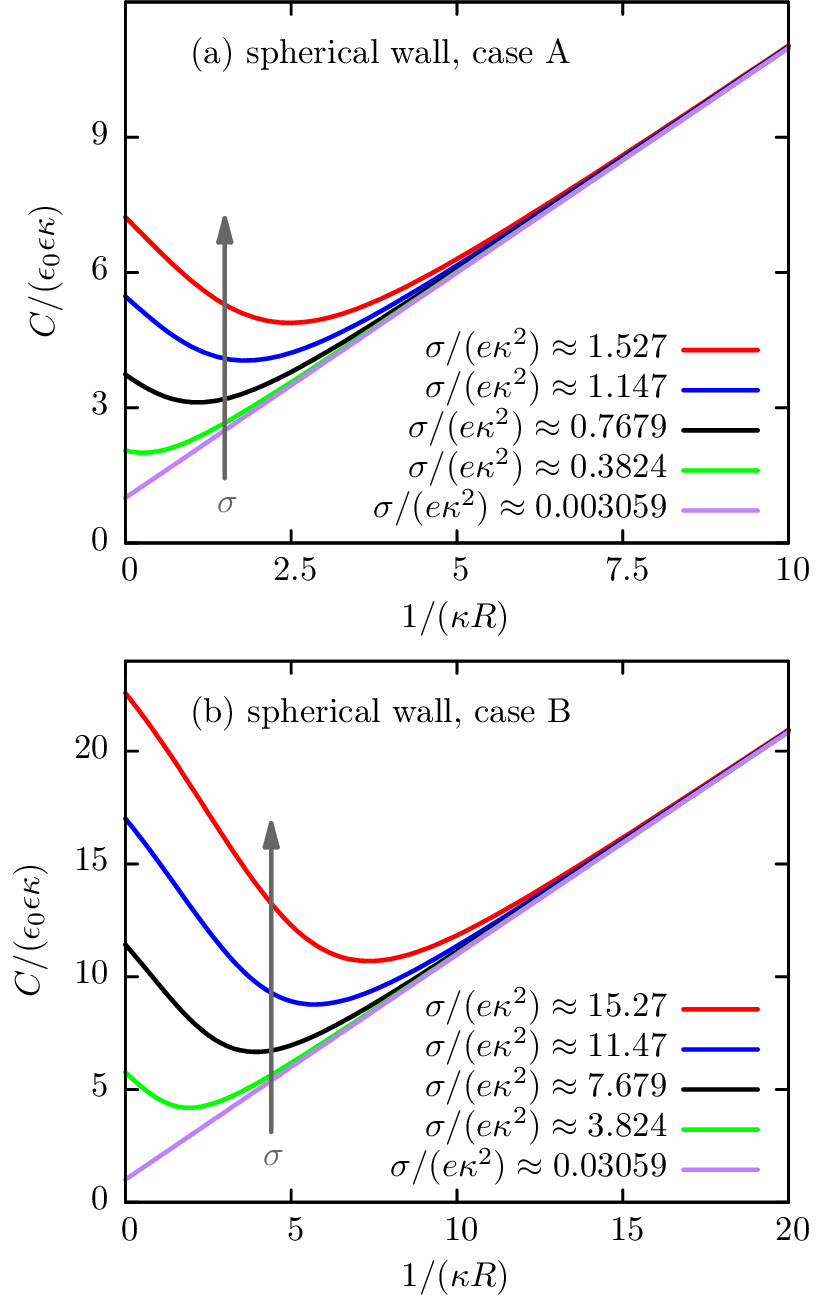}
   \caption{Reduced differential capacitance $C/(\epsilon_0\epsilon\kappa)$
            as a function of the dimensionless curvature $1/(\kappa R)$ of spherical electrodes. The data are obtained by solving
            the PB equation~(\ref{eq:PB_1D}) for
            two cases (A and B) of the bulk parameter choices (see the main text).
            Each curve corresponds to a constant value of the reduced surface
            charge density $\sigma/(e\kappa^2)$.
            The vertical arrow points in the direction of increasing $\sigma$.
           }
   \label{fig:PB_spherical}
\end{figure}

\begin{figure}[!t]
   \includegraphics[width=0.44\textwidth]{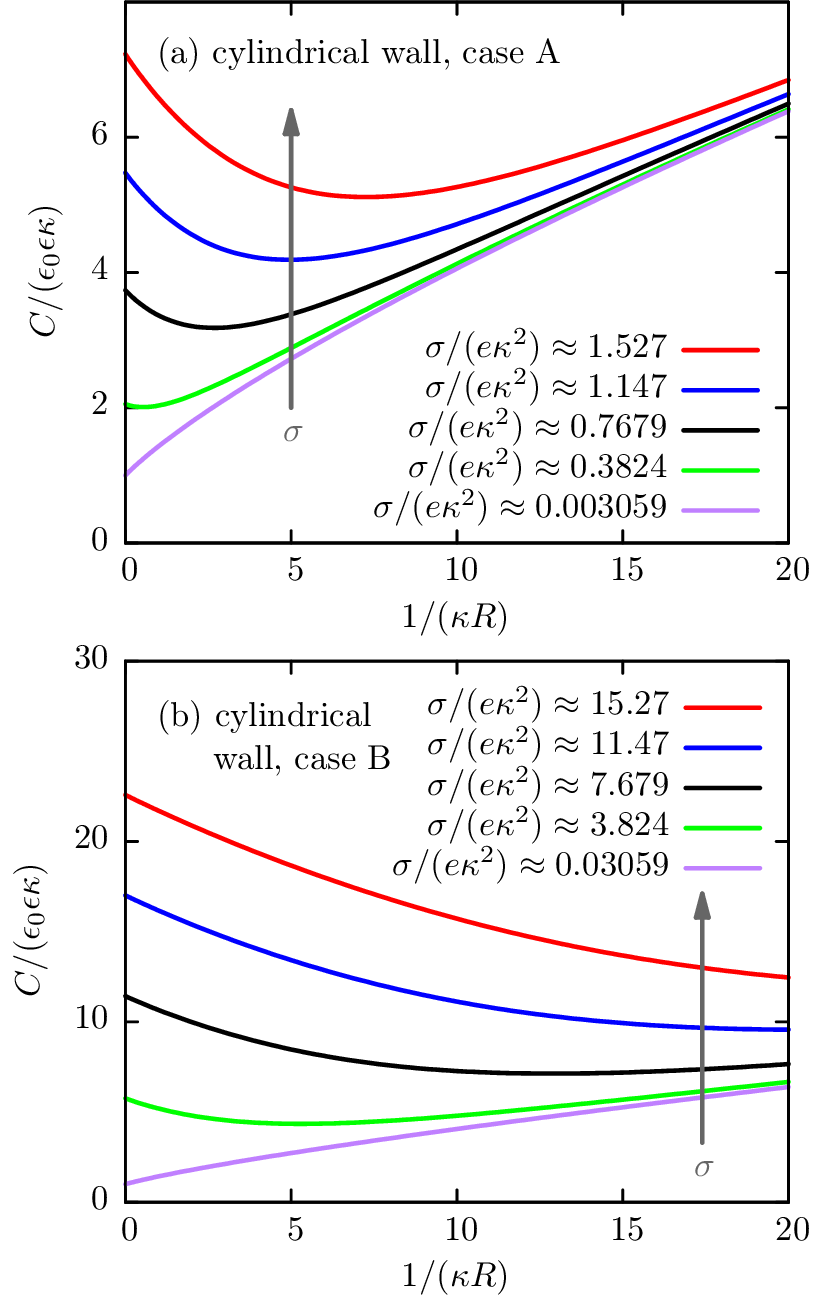}
   \caption{Same as Fig.~\ref{fig:PB_spherical} for cylindrical electrodes.}
   \label{fig:PB_cylindrical}
\end{figure}

The solution of the PB equation~(\ref{eq:PB_1D}) at the planar wall is available in closed form and the capacitance is given by \cite{Schmickler2010}
\begin{align}
   \frac{C}{\epsilon_0\epsilon\kappa}=\cosh\left[\frac{1}{2}\beta e\Phi(0)\right],
   \label{eq:C_planar}
\end{align}
where the potential at the wall $\Phi(0)$ depends on the surface charge density $\sigma$ as
\begin{align}
   \Phi(0)=\frac{2}{\beta e}\text{arsinh}\left(\frac{\beta e\sigma}{2\epsilon_0\epsilon\kappa}\right).
\end{align}
For spherical and cylindrical geometries the PB equation~(\ref{eq:PB_1D}) is solved numerically.
If lengths, charges, and energies are measured in units of the Debye length $1/\kappa$, the elementary charge $e$,
and the thermal energy $1/\beta=k_BT$, respectively,
the present model of a monovalent salt solution is specified by the following three dimensionless, independent parameters: $I/\kappa^3$,
$\kappa R$, and $\sigma/(e\kappa^2)$. In Figs.~\ref{fig:PB_spherical} and \ref{fig:PB_cylindrical}
results for the reduced capacitance are shown for two cases~A and B corresponding to
the choices
$I/\kappa^3\approx0.05329$ 
and $I/\kappa^3\approx0.1685$, respectively.
The parameters would, for example, refer to an aqueous electrolyte solution at room temperature $T=300\,\text{K}$
with relative permittivity $\epsilon=77.7003$, Debye length $1/\kappa\approx9.600\,\mathring{\text{A}}$
[$30.36\,\mathring{\text{A}}$], and ionic strength $I=0.1\,\text{M}$ [$0.01\,\text{M}$] in case~A [B].
In Figs.~\ref{fig:PB_spherical} and \ref{fig:PB_cylindrical}
the reduced capacitance $C/(\epsilon_0\epsilon\kappa)$ is plotted as function of the
dimensionless wall curvature $1/(\kappa R)$ for various values of the reduced surface charge density $\sigma/(e\kappa^2)$. The value
$1/(\kappa R)=0$ corresponds to the planar wall result in Eq.~(\ref{eq:C_planar}). Larger values on the
horizontal axes are equivalent to larger curvatures and hence to smaller radii of the wall. Since the PB equation originates from a classic theory the results for
large curvatures should
be treated with caution.
In case~A, for example, the Debye length is about $1/\kappa\approx10\,\mathring{\text{A}}$.
This means that for $1/(\kappa R)>10$ the wall radius is smaller than the atomic length scale of $1\,\mathring{\text{A}}$.
Within this range the particle size, which is not captured by the PB theory, should play a role.
In the case of the spherical wall (Fig.~\ref{fig:PB_spherical}) the curvature dependence of the capacitance
for the smallest chosen value of $\sigma$ almost coincides with a straight line.
This is in accordance with the analytic result in Eq.~(\ref{eq:C_sph_lin}) which renders a linear polynomial in $1/(\kappa R)$.
For increasing $\sigma$ the capacitance in the planar limit $1/(\kappa R)=0$ increases, whereas for large curvatures
the curves seem to converge from above to the graph for
$\sigma\rightarrow0$. This indicates that the linear theory becomes the more valid the larger the curvature is chosen.
In between the limits of high and low curvatures the capacitance exhibits a minimum the position of which shifts with $\sigma$.
Also the slope of the graph for small curvatures depends on $\sigma$. For $\sigma\rightarrow0$ there is a positive slope,
whereas larger surface charge densities give rise to a negative slope.

The curvature dependences in Fig.~\ref{fig:PB_cylindrical} of the capacitances for electrolytes in contact with
cylindrical walls resemble the ones at spherical walls in Fig.~\ref{fig:PB_spherical}. The
results for cylinders look like the results for spheres stretched in horizontal direction.
However, for $\sigma\rightarrow0$ the capacitance at a cylindrical wall clearly deviates from a linear function [see Eq.~(\ref{eq:C_cyl_lin})].
The potential $\Phi(R)$ at the cylindrical electrode agrees well with the expression for
the surface potential in Ref.~\cite{Trizac2007} [Eqs.~(3) and (4) therein] within the specified range of validity,
i.e., for not too small curvatures and line charge densities.

The linearized PB equation~(\ref{eq:lPB}) corresponds to the lowest curves in Figs.~\ref{fig:PB_spherical} and \ref{fig:PB_cylindrical}.
Thus important features, particularly in the range of small curvatures, are neglected, whereas for large curvatures a description
based on the linear theory appears to be sufficient.
In the solution of the full equation~(\ref{eq:PB_1D}) the surface charge density affects the capacitance for small curvatures to a large extent whereas for large
curvatures the behavior becomes more and more general and independent of $\sigma$. This phenomenon will be addressed in the following sections.

\subsection{\label{subsec:large_radii}Limit of large wall radii}
Within this subsection we focus on walls with large radii $\kappa R\gg1$ or small curvatures $1/(\kappa R)\ll1$.
It has been shown before that in this limit the capacitance as function of the curvature varies strongly
with the surface charge density $\sigma$ (see Figs.~\ref{fig:PB_spherical} and \ref{fig:PB_cylindrical}
and the discussion in the previous Sec.~\ref{subsec:PB}). In order to examine this observation in more detail,
the capacitance is taken as a power series in terms of small curvatures $(\kappa R)^{-1}\ll1$:
\begin{align}
   C=\epsilon_0\epsilon\kappa\sum\limits_{n=0}^\infty\frac{C_n}{(\kappa R)^n},
   \label{eq:C_curv_exp}
\end{align}
where $\epsilon_0\epsilon\kappa C_0$ is the capacitance of a planar wall [Eq.~(\ref{eq:C_planar})].
In Appendix~\ref{app:large_radii} the calculation of the dimensionless coefficients $C_n$ is explained in detail.
\begin{figure}[!t]
   \includegraphics[width=0.44\textwidth]{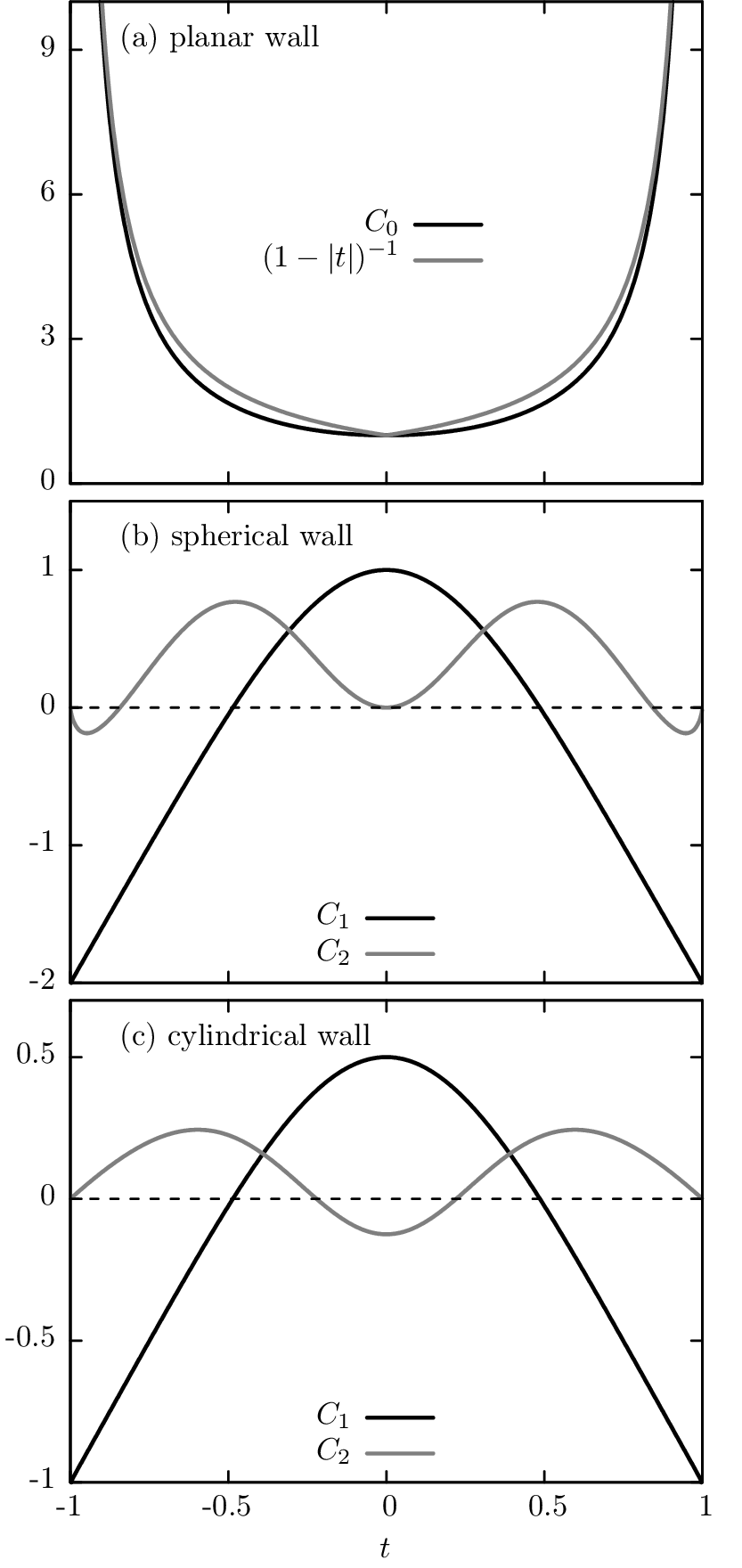}
   \caption{Lowest order coefficients $C_{0,1,2}$ of the curvature expansion [Eq.~(\ref{eq:C_curv_exp})] as function
            of the parameter $t$ defined in Eqs.~(\ref{eq:t}) and (\ref{eq:y0}). The entire information about the influences of $T,\epsilon,I$,
            and $\sigma$ is contained in the dependence on
            $t\in[-1,1]$ (see the discussion in Appendix~\ref{app:large_radii}). For clarity the coefficient $C_0$,
            describing the contribution of the planar wall to $C$
            [panel (a)], is displayed separately from the coefficients $C_1$ and $C_2$ for spherical (b) and cylindrical (c) walls.
           }
   \label{fig:PBexp}
\end{figure}
In Fig.~\ref{fig:PBexp} the lowest order coefficients $C_{0,1,2}$ of the curvature expansion in Eq.~(\ref{eq:C_curv_exp})
are plotted as function of the dimensionless parameter
\begin{align}
   t:=\tanh\left[\frac{1}{2}\text{arsinh}\left(\frac{\beta e\sigma}{2\epsilon_0\epsilon\kappa}\right)\right]\in[-1,1],
   \label{eq:t}
\end{align}
which is a combination of $T,\epsilon,I$, and $\sigma$ such that the sign of $t$ agrees with the sign of $\sigma$ [see also Eq.~(\ref{eq:y0})].
Apart from the geometry captured by $d$, the coefficients $C_n$ depend only on $t$. Thus within PB theory every parameter choice can be assigned to
Fig.~\ref{fig:PBexp}. Since the solutions $\Phi(r)$ of Eqs.~(\ref{eq:PB_1D}) and (\ref{eq:BC}) are odd functions of $\sigma$,
the capacitance in Eq.~(\ref{eq:C})
and hence the coefficients $C_n$ are even functions of $t$, i.e., $C_n(t)=C_n(-t)$; it is therefore sufficient
to only discuss the range $t\geq0$. For the planar wall [Fig.~\ref{fig:PBexp}(a)] the
formulation in terms of scaled variables is
\begin{align}
   C_0=\frac{1+t^2}{1-t^2}
\end{align}
which diverges for $\sigma\rightarrow\pm\infty\Leftrightarrow t\rightarrow\pm1$ as $(1-|t|)^{-1}$. The coefficient $C_1$
[Figs.~\ref{fig:PBexp}(b) and (c)]
exhibits the same qualitative behavior
for both curved walls: its value in the spherical case is twice of that in the cylindrical case.
At $t=0$, $C_1$ attains a positive maximum; 
for $t>0$ the curve decreases monotonically and crosses the $t$-axis at $t=0.4858(3)$. In Figs.~\ref{fig:PB_spherical} and \ref{fig:PB_cylindrical},
$C_1$ corresponds to the slope for small curvatures. Indeed the slope changes from
positive to negative with increasing $\sigma$, i.e., increasing $t$. The roots of $C_1$ in Figs.~\ref{fig:PBexp}(b) and (c) correspond to a special combination of parameters
for which the initial slope in Figs.~\ref{fig:PB_spherical} and \ref{fig:PB_cylindrical} would
be exactly zero. For $C_2$ qualitative differences between the curved wall shapes occur. In case of the spherical wall [Fig.~\ref{fig:PBexp}(b)]
$C_2$ is zero at $t=0$ which is consistent with
Fig.~\ref{fig:PB_spherical} showing a straight line for $\sigma\rightarrow0$. Increasing $t$ leads to a somewhat oscillatory behavior of $C_2$.
Positive values of $C_2$ correspond to a convex function (from below) in Fig.~\ref{fig:PB_spherical} for small curvatures
and intermediate values of $\sigma$
whereas negative values of $C_2$ for $t\rightarrow1$ indicate a concave behavior. The magnitude of negative values of $C_2$ is relatively
small so that the concave behavior
is less pronounced. However, the latter is visible in Fig.~\ref{fig:PB_spherical}(b) for large $\sigma$.
For the crossover value for $t$ between convex and concave we obtain $t=0.8428(3)$.
In the case of the cylindrical wall [Fig.~\ref{fig:PBexp}(c)] $C_2$ is negative at $t=0$ and consequently
in Fig.~\ref{fig:PB_cylindrical} the concave behavior for small curvatures and small
$\sigma$ is visible. Upon increasing $t$ the
coefficient $C_2$ changes sign from negative to positive and remains positive for values of $t$ larger than the
root at $t=0.2208(3)$. In Fig.~\ref{fig:PB_cylindrical} the convex
behavior can be observed for large $\sigma$.
This analysis shows that even for very large $\sigma$ no concave behavior can be expected as in the case of spherical walls. The coefficients
[see Eqs.~(\ref{eq:C_sph_lin}) and (\ref{eq:C_cyl_lin})]
obtained within the linearized theory [Eq.~(\ref{eq:lPB})] are covered by the present analysis and correspond
to the values at $t=0\Leftrightarrow\sigma=0$.

At this stage an excursion to \textit{morphometric thermodynamics} (MT) is appropriate.
Within that approach the interfacial tension $\gamma$ takes a very simple form with respect to the dependence on the geometry of the surrounding walls.
For the geometries of the current study the dependences on the radius $R$ can be formulated as
\begin{align}
   \begin{aligned}
      \gamma=
      \begin{cases}
         \displaystyle\gamma_0+\frac{\gamma^{s}_1}{R}+\frac{\gamma^{s}_2}{R^2}&,\text{spherical wall},\\[6pt]
         \displaystyle\gamma_0+\frac{\gamma^{c}_1}{R}&,\text{cylindrical wall}.
      \end{cases}      
   \end{aligned}
   \label{eq:gamma_morph}
\end{align}
Within MT there are no higher order terms and the coefficients $\gamma_n^{c,s}$ are independent of the radius $R$.
(For further details see Ref.~\cite{Koenig2004}.)
The connection with the present study is given by the Lippmann equation \cite{Schmickler2010}
\begin{align}
   \sigma=-\frac{\partial\gamma}{\partial\Phi(r_w)}
\end{align}
and hence
\begin{align}
   C=-\frac{\partial^2\gamma}{\partial\Phi(r_w)^2}.
   \label{eq:Lippmann2}
\end{align}
Since differentiation of Eq.~(\ref{eq:gamma_morph}) with respect to the electrode potential $\Phi(r_w)$ does not change the form of the equations, MT predicts
the same truncated curvature dependence for the capacitance $C$. However, examinations in terms of the capacitance have the advantage that this
quantity is uniquely defined contrary to the interfacial tension. In a previous study \cite{Reindl2015} we examined the implications of
various interface conventions concerning the accuracy of
MT in terms of the interfacial tension. We found that the quality of the approach as an approximation depends to a large extent on
the interfacial position which in principle may be chosen arbitrarily.
Following the prediction of MT in the case of cylindrical walls the coefficient $C_2$ should be zero.
However, in agreement with earlier work on curved interfaces using gradient expansion approaches \cite{Gompper1992,Blokhuis1993},
already linear theory [Eq.~(\ref{eq:C_cyl_lin})]
exhibits a nonzero coefficient and the full solution Fig.~\ref{fig:PBexp}(c) reveals that $C_2$ is nonzero for most choices of the parameters.
Therefore MT is not an exact approach, which
is not surprising because even for simple fluids its precision has been doubted recently
(see, e.g., Refs.~\cite{Reindl2015,Blokhuis2013,Urrutia2014,Goos2014}).
Therefore, MT has the status of an approximation. For example, when discussing cylindrical walls
the necessary restriction $|C_2|\ll|C_1|$  might be adequate to truncate the curvature expansion in accordance with MT
which is the case for values of $t$ far away from the root of $C_1$.
It is remarkable that this is the case for $|t|\rightarrow1\Leftrightarrow|\sigma|\rightarrow\infty$, i.e., for highly charged electrodes.
In any case this quality criterion depends on $t$ and therefore on the surface charge density. In general the curvature coefficients are properties of the fluid and the
wall-fluid interaction \cite{Koenig2004}. As a consequence, for simple fluids, the coefficients are fixed once a certain wall-fluid system has been chosen.
However, in the case of electrode-electrolyte systems, the wall-fluid interaction is typically not fixed but can be adjusted via the surface charge density.
For such cases the dependence of the coefficients $C_n$ on $\sigma$ has to be known.
This further complicates and reduces the applicability of MT.

\subsection{\label{subsec:small_radii}Limit of small wall radii}
For spherical walls and large curvatures a somewhat general behavior is observed (Fig.~\ref{fig:PB_spherical}):
all curves shown approach the straight line which corresponds to the results for small $\sigma$ and
which is in accordance with the result of the linearized theory [Eq.~(\ref{eq:C_sph_lin})]. In the case of cylindrical walls
(Fig.~\ref{fig:PB_cylindrical}) a similar behavior is visible; however,
the degree of convergence towards the curve corresponding to small $\sigma$ is inferior to that for
spherical walls, at least within the shown curvature interval.

Indeed, it is possible to show
analytically that non-linear contributions to the solution of the PB equation [Eq.~(\ref{eq:PB_1D})]
are negligible for sufficiently large curvature, e.g., if
\begin{align}
   \quad\frac{1}{\kappa R}\gg\sqrt{\frac{1}{\sqrt{6}}\frac{\beta e|\sigma|}{\epsilon_0\epsilon\kappa}}
   \label{eq:estimate}
\end{align}
for a spherical wall (see Appendix~\ref{app:small_radii} for details). This finding explains the general behavior
encountered in Fig.~\ref{fig:PB_spherical} because for any finite $\sigma$ there is a range
of (large) curvatures for which the inequality in Eq.~(\ref{eq:estimate}) holds. Cylindrical electrodes (Fig.~\ref{fig:PB_cylindrical}) exhibit
curvature dependent capacitances which resemble the spherical results (Fig.~\ref{fig:PB_spherical}), stretched in horizontal direction. 
This finding is supported by the linearized theory. In the limit of small radii $\kappa R\rightarrow0$ the strengths $|\Phi(R)|$ of the electrode potentials 
[Eqs.~(\ref{eq:PhiR_sph_lin}) and (\ref{eq:PhiR_cyl_lin})] are monotonically increasing functions of $R$ and the one at the cylindrical electrode
with the same $R$ is larger than the corresponding one at the spherical wall. On one hand this means that in the case of cylindrical walls
smaller radii or larger curvatures
are necessary in order to get the same value of $\Phi(R)$ as in the spherical case; this also holds for the capacitance.
On the other hand the linearized theory is based on small values of the dimensionless
potential $\beta e|\Phi|\ll1$. Thus, in the case of cylindrical walls the linearized theory turns into a reliable
description at smaller radii or larger curvatures
as compared to the spherical wall. From the comparison of Figs.~\ref{fig:PB_spherical} and \ref{fig:PB_cylindrical} it follows that a corresponding
estimate like the one in Eq.~(\ref{eq:estimate}) would lead to wall radii below molecular sizes (see the discussion in Sec.~\ref{subsec:PB})
and which therefore would be of no practical use.

\section{\label{sec:summary}Summary and outlook}
In terms of the Poisson-Boltzmann (PB) equation [Eq.~(\ref{eq:PB})] we have analyzed electrolytes in contact with electrodes of planar $(d=0)$,
cylindrical $(d=1)$, or spherical $(d=2)$ shape. The differential capacitance $C$ [Eq.~(\ref{eq:C})]
was calculated for various ionic strengths $I$, surface charge densities
$\sigma$, and electrode radii $R$. The focus was on examining the dependence of the capacitance on the curvature $1/R$ of the electrode as displayed in
Figs.~\ref{fig:PB_spherical} and \ref{fig:PB_cylindrical}.
In all cases the surface charge density has a strong effect on the capacitance for small curvatures whereas for large curvatures the behavior becomes 
independent of $\sigma$. These limits have been analyzed in detail. For small curvatures (see Sec.~\ref{subsec:large_radii}) we found that a curvature
expansion of the capacitance [Eq.~(\ref{eq:C_curv_exp})] reveals the behavior in a very convenient way because the corresponding expansion
coefficients $C_n$ depend on the single parameter $t\in[-1,1]$ [Eqs. (\ref{eq:t}) and (\ref{eq:y0})] and on the geometry $d\in\{0,1,2\}$ only.
Therefore, within PB theory, the influence of any conceivable combination of system parameters on the lowest order coefficients $C_n$ can be
inferred from Fig.~\ref{fig:PBexp}. For large curvatures (see Sec.~\ref{subsec:small_radii}) an analytic discussion provides the insight that the
linearized PB theory becomes reliable, if the radius of
the spherical wall is chosen to be small enough; this explains the general behavior visible in Fig.~\ref{fig:PB_spherical}.

In the present study the mesoscopic structure of electrolyte solutions at curved electrodes has been discussed systematically in terms of the capacitance
within PB theory (i) because this approach is widely used in various research fields, and (ii) because it offers to judge less integral,
microscopic approaches such as the one presented in part II of this study \cite{Reindl2016}.

%-------------------------------------------------------------------------------

\appendix

\section{\label{app:large_radii}Limit of large wall radii}

We assume that for large radii $R\rightarrow\infty$ the solution of the PB equation~(\ref{eq:PB_1D}) can be expanded in terms of
powers of the curvature such that the dimensionless potential $y:=\beta e\Phi$ takes the form
\begin{align}
   y(r=R+z)=\sum\limits_{n=0}^\infty \frac{y_n(z)}{(\kappa R)^n},
   \label{eq:y_curv_exp}
\end{align}
where $z\in[0,\infty)$ measures the distance from the wall. For all radii the boundary conditions in Eq.~(\ref{eq:BC})
translate into an inhomogeneous condition at the wall $z=0$,
\begin{align}
   \begin{aligned}
      y_0'(z)\Big|_{z=0}=-\frac{\beta e\sigma}{\epsilon_0\epsilon},\quad
      y_{n>0}'(z)\Big|_{z=0}=0,
   \end{aligned}
\end{align}
and a homogeneous one at $z=\infty$:
\begin{align}
   y_{n\geq0}(\infty)&=0.
\end{align}
The ansatz in Eq.~(\ref{eq:y_curv_exp}) gives rise to a curvature expansion of the PB equation~(\ref{eq:PB_1D}).
In the following, we take into account orders up to and including $1/(\kappa R)^2$. The lowest order leads to
\begin{align}
   y_0''(z)=\kappa^2\sinh[y_0(z)],
   \label{eq:PB_planar}
\end{align}
which is the PB equation~(\ref{eq:PB_1D}) with $d=0$ for the dimensionless potential $y_0(z)$ at the planar wall.
The orders which are linear and quadratic in the curvature $1/(\kappa R)$ correspond to differential equations
for the spatially varying expansion coefficients $y_{1,2}(z)$:
\begin{align}
   \begin{aligned}
      \frac{y_n''(z)}{\kappa^2}&-\cosh[y_0(z)]y_n(z)=\\
      &=
      \begin{cases}
         n=1:& \displaystyle -d\frac{y_0'(z)}{\kappa},\\[6pt]
         n=2:& \displaystyle -d\frac{y_1'(z)}{\kappa}+zdy_0'(z)\\[6pt]
             & \displaystyle +\frac{1}{2}\sinh[y_0(z)]y_1(z)^2,
      \end{cases}\\
   &d\in\{1,2\},\quad n\in\{1,2\}.
   \end{aligned}
   \label{eq:DEQ_y12}
\end{align}
Within the curvature expansion given by Eq.~(\ref{eq:y_curv_exp}) the contribution of the planar wall is entirely captured by the coefficient $y_0$.
Higher order coefficients $y_{n>0}$ occur solely at curved walls $(d\neq0)$.
The solution of Eq.~(\ref{eq:PB_planar}) is given by (see Ref.~\cite{Schmickler2010})
\begin{align}
   \begin{aligned}
      y_0(z)&=4\,\text{artanh}[t\exp(-\kappa z)],\\
      t&:=\tanh\left[\frac{1}{2}\text{arsinh}\left(\frac{\beta e\sigma}{2\epsilon_0\epsilon\kappa}\right)\right]\\
       &=\frac{2\epsilon_0\epsilon\kappa}{\beta e\sigma}\left[\sqrt{1+\left(\frac{\beta e\sigma}{2\epsilon_0\epsilon\kappa}\right)^2}-1\right]\in[-1,1].
   \end{aligned}
   \label{eq:y0}
\end{align}
In Eq.~(\ref{eq:y0}) the dependence on $z$ maps onto
the scaled spatial variable
\begin{align}
   \begin{aligned}
      x&:=t\exp(-\kappa z),\quad|x|\in[0,|t|],\;\text{with}\\
      f_n(x)&:=y_n(z(x)),
   \end{aligned}
   \label{eq:x}
\end{align}
such that, e.g., the planar wall result takes the simple form
\begin{align}
   f_0(x)=4\,\text{artanh}(x).
   \label{eq:f0}
\end{align}
The differential equations for $f_{1,2}(x)$ are given by
\begin{align}
   \begin{aligned}
      x^2f''_n&(x)+xf'_n(x)-\cosh[f_0(x)]f_n(x)=\\
      &=\begin{cases}
          n=1:&xdf'_0(x),\\[6pt]
          n=2:&\displaystyle xdf_1'(x)+d\ln\left(\frac{x}{t}\right)xf'_0(x)\\[6pt]
              &\displaystyle +\frac{1}{2}\sinh[f_0(x)]f_1(x)^2,
       \end{cases}\\
       &d\in\{1,2\},\quad n\in\{1,2\},
   \end{aligned}
   \label{eq:DEQ_f12}
\end{align}
subject to the boundary conditions
\begin{align}
   \begin{aligned}
      f_{1,2}(x=0)=0,\\
      f'_{1,2}(x)\Big|_{x=t}=0.
   \end{aligned}
   \label{eq:BC_f12}
\end{align}
From Eqs.~(\ref{eq:DEQ_f12}) and (\ref{eq:BC_f12}) it follows that the scaled potentials $f_{1,2}$ depend parametrically on $d\in\{1,2\}$ and $t$.
Moreover, the differential equations~(\ref{eq:DEQ_f12}) are to be solved within a finite domain of values of $x$ [Eq.~(\ref{eq:x})]
so that the whole parameter space can be scanned rapidly. Finally, the capacitance follows as
\begin{align}
   \begin{aligned}
      C&=\left[\frac{\partial\Phi(r_w)}{\partial\sigma}\right]^{-1}=\left[\frac{1}{\beta e}\frac{\partial}{\partial\sigma}\sum\limits_n\frac{y_n(z=0)}{(\kappa R)^n}\right]^{-1}\\
       &=\left[\frac{1}{\beta e}\frac{\partial}{\partial\sigma}\sum\limits_n\frac{f_n(x=t)}{(\kappa R)^n}\right]^{-1}\\
       &=\epsilon_0\epsilon\kappa\left[\sum\limits_n\frac{1}{4}\frac{(1-t^2)^2}{1+t^2}\frac{\partial f_n(x=t)}{\partial t}\frac{1}{(\kappa R)^n}\right]^{-1}\\
       &=:\epsilon_0\epsilon\kappa\sum\limits_n\frac{C_n}{(\kappa R)^n},
   \end{aligned}
   \label{eq:Cn}
\end{align}
which defines the dimensionless expansion coefficients $C_n$ of the differential capacitance $C$.
Alternatively, the coefficients $C_{1,2}$ can be determined from the expressions for the surface potential
of spherical and cylindrical surfaces which are given in Ref.~\cite{Lekkerkerker1989}. In Eq.~(\ref{eq:Cn}) the
expression $\partial f_n(x=t)/\partial t$ refers to the derivative of $f_n$ with respect to $t$ after
evaluation at $x=t$; the dependence on $d$ and $t$ is transferred to the coefficients $C_n$. 
Apart from the influence of the geometry via $d$, the whole parameter space given by $T,\epsilon,I$, and $\sigma$ is contained in $t\in[-1,1]$.
Thus every parameter choice within PB theory can be assigned to Fig.~\ref{fig:PBexp}.

\section{\label{app:small_radii}Limit of small wall radii}
Here we investigate under which conditions the non-linearities of the full PB equation~(\ref{eq:PB_1D}) may be neglected.
To this end, with the dimensionless potential $y:=\beta e\Phi$ we consider the equation
\begin{align}
   \begin{aligned}
      \frac{1}{r}\frac{\partial^2}{\partial r^2}&\left[r\lambda y(r)\right]=\kappa^2\sinh[\lambda y(r)]\\
      &=\kappa^2\left[\lambda y(r)+\frac{1}{6}\lambda^3y(r)^3+O\left(\lambda^5\right)\right]
   \end{aligned}
   \label{eq:truncated_PB}
\end{align}
in spherical geometry, where $\lambda\in\mathbb{C}$ is an arbitrary complex parameter. For $\lambda=1$ the PB equation~(\ref{eq:PB_1D}) is recovered, whereas for $|\lambda|\ll1$
non-linearities on the right-hand side are suppressed.
In order to solve the truncated version of Eq.~(\ref{eq:truncated_PB}) we make the ansatz
\begin{align}
   y(r)=y_0(r)+\lambda^2y_2(r)+O\left(\lambda^4\right).
\end{align}
In lowest order $O\left(\lambda^0\right)$ the linearized PB equation~(\ref{eq:lPB}) with $d=2$ is recovered for which the spatially varying potential $y_0(r)$
and the electrode potential $y_0(R)$ are given by
\begin{align}
   \begin{aligned}
      &y_0(r)=A\frac{\exp(-\kappa r)}{r},\\
      &\quad\quad A:=\frac{sR^2}{1+\kappa R}\exp(\kappa R),\quad\quad s:=\frac{\beta e\sigma}{\epsilon_0\epsilon},\;\text{and}\\
      &y_0(R)=\frac{sR}{1+\kappa R},
   \end{aligned}
   \label{eq:y0_lin}
\end{align}
respectively.
The next higher order $O\left(\lambda^2\right)$ leads to a differential equation for the dominant non-linear contribution $y_2(r)$
\begin{align}
   \frac{1}{r}\frac{\partial^2}{\partial r^2}&\left[r y_2(r)\right]=\kappa^2 y_2(r)+\frac{\kappa^2}{6}y_0(r)^3,
   \label{eq:DEQ_y2}
\end{align}
where the inhomogeneity is given by the solution of the linearized PB equation [Eq.~(\ref{eq:y0_lin})].
The solution of Eq.~(\ref{eq:DEQ_y2}) takes the form
\begin{align}
   \begin{aligned}
      y_2(r)&=B\frac{\exp(-\kappa r)}{r}+\frac{f(r)}{r},\;\text{where}\\
      f(r)&:=-\frac{\kappa A^3}{12}\int\limits_R^\infty \ud r'\,\exp(-\kappa|r-r'|)\frac{\exp(-3\kappa r')}{r'^2},\\
      B&:=f(R)\frac{\kappa R-1}{\kappa R+1}\exp(\kappa R).
   \end{aligned}
\end{align}
The electrode potential $y_2(R)$ is given by
\begin{align}
   \begin{aligned}
      y_2(R)&=-y_0(R)\frac{1}{6}\left(\frac{s}{\kappa}\right)^2\frac{h(4\kappa R)}{\frac{1}{\kappa R}\left(1+\frac{1}{\kappa R}\right)^3},\;\text{with}\\
      h(z)&:=z\exp(z)\int\limits_z^\infty \ud x\,\frac{\exp(-x)}{x^2}.
   \end{aligned}
   \label{eq:y2}
\end{align}
From these results one infers that the contribution from the linearized PB equation is the dominant one if
the leading non-linear term $y_2(R)$ is much smaller than the linear one $y_0(R)$. Since $h(z>0)\leq1$, one obtains
\begin{align}
   \begin{aligned}
      \left|\frac{y_2(R)}{y_0(R)}\right|&\ll1\\
      &\text{for}\quad\frac{1}{\kappa R}\gg\sqrt{\frac{1}{\sqrt{6}}\frac{|s|}{\kappa}}=\sqrt{\frac{1}{\sqrt{6}}\frac{\beta e|\sigma|}{\epsilon_0\epsilon\kappa}}.
   \end{aligned}
\end{align}
Hence for any finite $\sigma$ the first non-linear term $y_2(R)$ becomes negligible if the curvature
$(\kappa R)^{-1}$ is chosen large enough. For small radii $R$ the linearized PB theory turns into a reliable description.

%-------------------------------------------------------------------------------

\end{document}